\documentclass[aps,prb,twocolumn,showpacs,superscriptaddress,groupedaddress,reprint]{revtex4-1}
\usepackage{graphicx}
\usepackage{mathrsfs}
\usepackage{natbib}
\usepackage{color}
\usepackage{epstopdf}
\usepackage{cleveref}
\usepackage{amsmath}
\usepackage{tabularx}

\newcolumntype{C}{>{\centering\arraybackslash}X} 

\begin{document}

\widetext

\title {Diffusion in systems crowded by active force-dipole molecules}

\author{Matthew Dennison$^{1}$} \author{Raymond Kapral$^{1,2}$}
\author{Holger Stark$^{1}$}

\affiliation{$\,^{1}$ Institut f\"{u}r Theoretische Physik, Technische
Universit\"{a}t Berlin, Hardenbergstrasse 36, 10623 Berlin, Germany}
\affiliation{$\,^{2}$ Chemical Physics Theory Group, Department of
Chemistry, University of Toronto, Toronto, Ontario M5S 3H6, Canada}

\date{\today}

\begin{abstract}
  Experimental studies of systems containing active proteins that
  undergo conformational changes driven by catalytic chemical
  reactions have shown that the diffusion coefficients of passive
  tracer particles and active molecules are larger than the
  corresponding values when chemical activity is absent. Various
  mechanisms have been proposed for such behavior, including, among
  others, force dipole interactions of molecular motors moving on
  filaments and collective hydrodynamic effects arising from active
  proteins. Simulations of a multi-component system containing active
  dumbbell molecules that cycle between open and closed states, a
  passive tracer particle and solvent molecules are carried
  out. Consistent with experiments, it is shown that the diffusion
  coefficients of both passive particles and the dumbbells themselves
  are enhanced when the dumbbells are active. The dependence of the
  diffusion enhancement on the volume fraction of dumbbells is
  determined, and the effects of crowding by active dumbbell molecules
  are shown to differ from those due to inactive molecules.
\end{abstract}

\maketitle

\section{Introduction}\label{sec:intro}

A body of evidence points to the existence and importance of
nonthermal fluctuations in the cell that are driven by chemical
activity that maintains the cell in a nonequilibrium
state.~\cite{ref:Caspi,ref:Lau,ref:Fred_motor,ref:Wilhelm08,ref:Gallet,ref:Weber,ref:Bruinsma,ref:Guo}
Such fluctuations have been measured and characterized using various
experimental probes and are often attributed to the forces generated
by molecular motors when they interact with filaments comprising the
cellular cytoskeletal network.  The results of {\it in vitro}
experiments on systems containing actin networks and active myosin
motors also suggest that nonthermal fluctuations play a significant
role in the systems' dynamical response to
deformation.~\cite{ref:Mizuno,ref:Toyota} Support for such effects is
provided by the observation that the mean square displacements of
passive molecules and active species are smaller when the production
of ATP is inhibited.  For example, the diffusive dynamics of
chromosomal loci in prokaryotic cells is sensitive to metabolic
activity; when ATP synthesis is inhibited the apparent diffusion
coefficient decreases.~\cite{ref:Weber} Force-spectrum-microscopy
studies have shown that force fluctuations in eukaryotic cells enhance
the movement of large and small molecules; when the activity of myosin
II motors is selectively inhibited diffusive motion decreases but not
to the degree when all ATP synthesis is suppressed.~\cite{ref:Guo}
These studies have concluded that it is the aggregate of all metabolic
activity, and not just that of motor proteins, that contributes to
enhanced diffusive motion.

Enhanced diffusion of enzymes and passive particles has also been
observed in {\it in vitro} studies of active enzymes in solution where
motor proteins are not
present~\cite{ref:Sen-enz10,ref:Sengupta13,ref:Sen-grubbs}, and the
possible origins of these effects have been
discussed~\cite{ref:Golestanian}. Proteins executing conformational
changes as a result of catalytic chemical activity can give rise to
collective hydrodynamic effects that enhance the diffusion of both
passive particles and
enzymes.~\cite{ref:Mikhailov15,Kapral16,ref:Mikhailov16} On more
macroscopic scales, and in a somewhat different context, experimental
and theoretical investigations of the diffusion coefficients of
passive particles in suspensions of active microorganisms have shown
diffusion enhancement due to the hydrodynamic flow fields generated by
their swimming
motions.~\cite{Kim_Breuer_04,Leptos_09,Mino_13,saintillan2012,Kasyap_14}

In this paper we investigate diffusive dynamics in a system containing
active dumbbell molecules, a passive particle and solvent. The active
dumbbell molecules cycle between open and closed conformations and act
as nonequilibrium fluctuating force dipoles. The microscopic dynamics
accounts for direct hydrodynamic interactions as well as direct
interactions among the dumbbell molecules. Molecular crowding is known
to influence the diffusive properties of tracer particles in solutions
where the concentration of crowding species is high: subdiffusive
dynamics is observed on intermediate time scales and long-time
diffusion coefficients decrease as the concentration of crowding
elements
increases.~\cite{Metzler00,Schnell04,HF13,Nakano14,Kuznetsova14} Our
investigations show how the diffusive dynamics of passive particles,
and the dumbbells themselves, vary with the volume fraction of active
dumbbell molecules. Comparisons with the results for the diffusive
dynamics in systems containing inactive dumbbells allow us to analyze
and describe the effects of dumbbell activity.

The outline of the paper is as follows. Section~\ref{sec:method}
describes the system under study, including the active dumbbell-shaped
``molecules", the interaction potentials among species and the
dynamical method used to evolve the system. The properties of single
active and inactive dumbbell molecules are presented in
Sec.~\ref{sec:single}. Systems containing many active dumbbell
molecules are considered in Sec.~\ref{sec:diffusion}, where simulation
results for the self-diffusion coefficients of the passive particle
and dumbbell molecules are given as a function of the dumbbell force
constants and volume fractions. The conclusions of the study are given
in Sec.~\ref{sec:Conclusions}.

\section{System and dynamical model}
\label{sec:method}

The entire system is comprised of dumbbell-shaped active molecules, a
passive particle and solvent molecules. The dumbbell-shaped molecules
consist of two beads linked by a harmonic bond, with interactions
between beads of different dumbbells described by the steep repulsive
potential function~\cite{ref:Padding},
\begin{equation}
\label{eq:Ucc}
U_{cc} = 4\epsilon\left[\left(\frac{\sigma_{c}}{r}\right)^{48}-\left(\frac{\sigma_{c}}{r}\right)^{24}+\frac{1}{4}
\right] \;\; r \le 2^{1/24}\sigma_{c},
\end{equation}
and zero otherwise, where $\epsilon$ sets the energy scale, which we
take to be $\epsilon=2.5 \; k_{B}T$ throughout, with $k_{B}$ the
Boltzmann constant and $T$ the temperature. The passive particle is a
structureless bead of diameter $\sigma_{c}$ and mass $m_{c}$, and
interacts with the dumbbell beads through the potential in
Eq.~(\ref{eq:Ucc}). The passive particle interacts with the solvent
particles through the repulsive Lennard-Jones interaction potential
$U_{cf}$, given by
\begin{equation}
\label{eq:Ucf}
U_{cf} = 4\epsilon\left[\left(\frac{\sigma_{cf}}{r}\right)^{12}-\left(\frac{\sigma_{cf}}{r}\right)^{6}+\frac{1}{4}
\right] \;\; r \le 2^{1/6}\sigma_{cf},
\end{equation}
and zero otherwise, where $\sigma_{cf}$ is the passive
particle-solvent interaction distance.

Other interactions involving the solvent are taken into account
through multiparticle collision (MPC) dynamics, comprising streaming
and collision steps.~\cite{ref:SRD,*ref:SRD2,Kapral_08,ref:Gompper}
The solvent molecules are represented by point particles of mass $m_f$
which are evolved in the streaming step, either ballistically or, when
potential interactions are present, by Newton's equations of
motion. In the collision steps, which occur at time intervals
$\tau_{{\text MPC}}$, the solvent particles are sorted into cubic
collision cells with length $a$, in which they interact with each
other according to multiparticle collisions. The coupling of the
dumbbell to the solvent also can be accounted for in this way, where
the constituent spheres of the dumbbell are included with the solvent
particles in the collision step, in the same manner as for single
polymers \cite{ref:polymer2,ref:polymer3}. Further simulation details
on the implementation of the MPC algorithm, along with parameter
values, are given in Appendix A.

\subsection*{Dumbbell molecule}

While the dumbbells are fictitious "molecules" their dynamics is
constructed to mimic the conformational changes that occur in active
enzymes.~\cite{kogler12,ref:Mikhailov15} Many catalytically active
proteins cycle between open and closed conformations: substrate
binding triggers passage from the open to closed state, while
substrate unbinding or product release causes the protein to return to
its open conformation. Such systems are maintained in a nonequilibrium
state by input of substrate and removal of product.

The dumbbell beads, each with mass $m_b$, are linked by a harmonic
bond that specifies open (large bond rest length) and closed (small
bond rest length) conformations. The
bond potential energy function has the form,
\begin{equation}
\label{eq:En_spring}
U=\displaystyle\frac{1}{2}k_0(t)\left(\ell-\ell_{0}(t)\right)^{2},
\end{equation}
where the bond rest length $\ell_0(t)$ and force constant $k_0(t)$ are
dichotomous random variables that take the two values
$\{\ell_o,\ell_c\}$ and $\{k_o,k_c\}$. These correspond to the values
for the open and closed configurations, respectively. A stochastic
process that switches the dumbbell between the open and closed states
is as follows: suppose the current rest bond length is $\ell_c$. If
during the evolution the bond length $\ell$ crosses a threshold and
satisfies the condition $\ell<\ell_{c}+\delta\ell_{c}$, a random time
$t_{h}$ is drawn from a log-normal distribution with average
$t_{c}$. The rest length and force constant will remain as
$\ell_{0}(t)=\ell_{c}$ and $k_{0}(t)=k_{c}$ for this time, after which
$\ell_0(t)$ is set to $\ell_{o}$ and the force constant to
$k_{0}(t)=k_{o}$. Similarly, if $\ell>\ell_{o}-\delta\ell_{o}$,
$\ell_0(t)$ is set to $\ell_{0}(t)=\ell_{c}$ and $k_0(t)$ to $k_c$
after a randomly chosen time $t_{h}$ with average $t_{o}$. This model
captures the gross features of active proteins that adopt open and
closed metastable conformations and operate though Michaelis-Menten
kinetics, $E+S\mathrel{ \mathop{\kern0pt
    {\rightleftharpoons}}\limits^{{\mathrm{k}_1}}_{\mathrm{k}_{-1}}} C
\stackrel{\mathrm{k}_{cat}}{\rightarrow } E+P$, where $E,\;S,\;C$ and
$P$ represent the enzyme, substrate, enzyme-substrate complex and
product, respectively, with excess substrate supplied and product
removed.

We shall call dumbbells that undergo such nonequilibrium cyclic
conformational changes {\it active} dumbbells. If instead the
stochastic mechanism responsible for these changes is absent and only
thermal fluctuations are present, the dumbbells will be termed {\it
  inactive} dumbbells. In our model, the inactive dumbbells will
simply fluctuate around the open conformation. This corresponds to a
system where enzymes are not supplied with substrate and remain in
open conformations.

\subsection*{Units and parameters}

Results are reported in dimensionless units: lengths are scaled by the
MPC cell size $a$, masses by the solvent particle mass $m_f$, energy
by $k_BT$ and time by $(a^2 m_f/k_BT)^{1/2}$. The spring constant $k$
is in units of $k_BT/a^2$.  In the simulations presented below we set
$t_{o}=0$ and vary $t_c$, the average time spent in the closed
conformation. Furthermore, we let $k_o=k$ and choose $k_c= 2k$ and
$\ell_c = \ell_o/2$.  Our choice of $t_{o}=0$ corresponds to a system
with excess substrate and reaction rates $\mathrm{k}_1, \;
\mathrm{k}_{cat} \gg \mathrm{k}_{-1}$. The closed and open dumbbell
bond lengths used in all of the simulation results are $\ell_{c}=2$
and $\ell_{o}=4$, respectively, and
$\delta\ell_{c}=\delta\ell_{o}=0.05( \ell_{o}-\ell_{c})=0.1$. Solvent
conditions will be indicated by the value of $\tau_{{\text
    MPC}}$. Unless stated otherwise, simulations use $\tau_{{\text
    MPC}}=0.01$, but some results will be presented for $\tau_{{\text
    MPC}}=0.05$ to explore the effects of different solvent conditions
(see Appendix A).

\section{Properties of single active and inactive dumbbells} \label{sec:single}

\begin{figure}[hbtp]
  \begin{center}
    \includegraphics[height=1.0\columnwidth,angle=270]{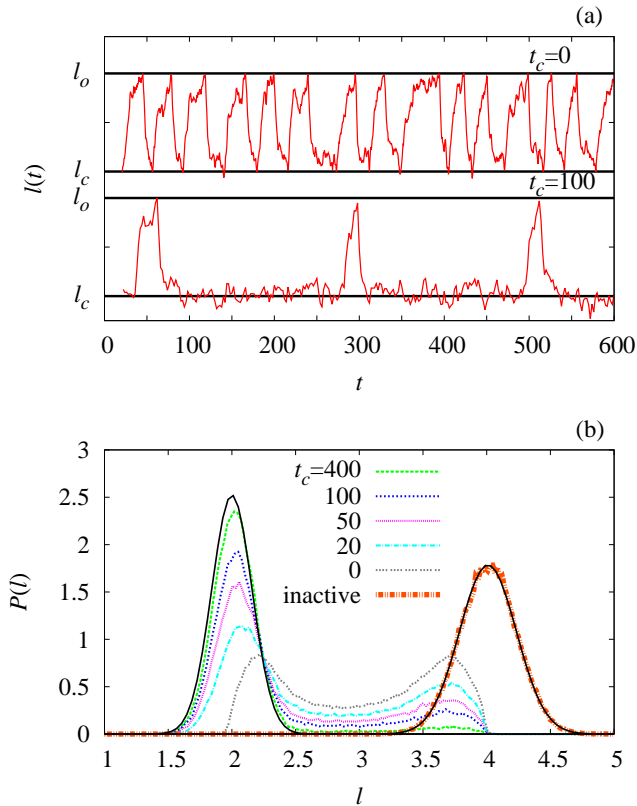}
    \caption{(a) Plot of the instantaneous bond length $\ell(t)$
      against time $t$ for an active dumbbell with force constant
      $k=20$. Results are presented for system with average hold times
      $t_{c}=0$ (top) and $t_{c}=100$ (bottom). (b) Probability
      density of lengths $P(\ell)$ against length $\ell$ for the above
      systems with various average hold times $t_{c}$ indicated in the
      plot; $P(\ell)$ for inactive dumbbells is also shown. The solid
      black line on the left peak shows a Gaussian distribution
      centred around the closed configuration, with mean $\ell_c$ and
      variance $\sigma^2=k_BT/k_c$, and the solid black line on the
      right peak shows a Gaussian distribution centred around the open
      configuration, with mean $\mu=\ell_o$ and variance
      $\sigma^2=k_BT/k_o$.}
    \label{fig:l_t}
  \end{center}
\end{figure}
Figure~\ref{fig:l_t}(a) shows how the bond length $\ell(t)$ of a
single active dumbbell varies with time as the dumbbell cycles between
open and closed conformations. The open and closed rest lengths are
indicated by the solid horizontal lines. Data for two values of the
average time spent in the closed conformation, $t_c=0$ and $t_c=100$,
are presented. When the dumbbell is in the metastable open or closed
states its dynamics will be controlled by thermal fluctuations about
the rest values $\ell_o$ and $\ell_c$ of these states. The average
time for a complete open-close cycle, $t_{\rm{cy}}$, is dominated by
$t_{c}$ when $t_{c}>t_{t}$, where $t_{t}$ is the time taken to pass
from one metastable state to the other. We note that $t_{t}$ will
depend on both the force constants $\{k_o,k_c\}$ and the rest bond
lengths $\{\ell_o,\ell_c\}$. In Fig.~\ref{fig:l_t}(b) we show the
probability density $P(\ell)$ of bond lengths for a range of average
hold times $t_{c}$. For $t_c=0$ we see two peaks of roughly equal size
centered near to but smaller than the open and closed configurations,
as the rest length switches between the two values $\{\ell_o =4,
\ell_c = 2\}$.  As we increase $t_c$ the peak close to $\ell_o$
decreases while the one about $\ell_c$ becomes more pronounced,
approaching that of a Gaussian distribution with mean $\mu=\ell_c$ and
variance $\sigma^2=k_BT/k_c$, resulting from thermal motion about
$\ell_c$. We also present data for an inactive dumbbell, corresponding
to a protein in the absence of substrate that remains in the open
configuration, which also exhibits a Gaussian distribution with mean
$\mu=\ell_o$ and variance $\sigma^2=k_BT/k_o$.

\begin{figure}[!ht]
  \begin{center}
    \includegraphics[height=1.0\columnwidth,angle=270]{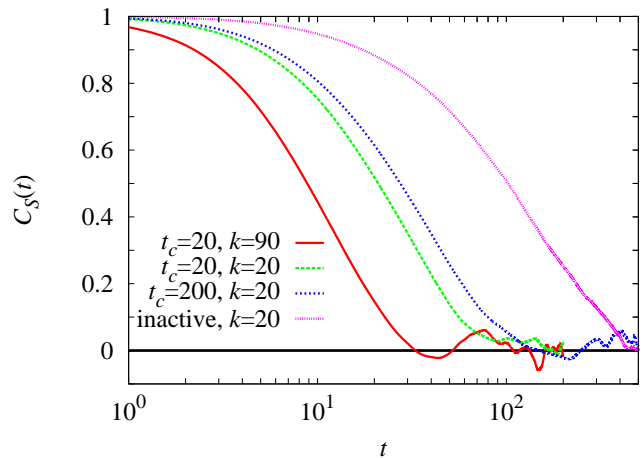}
    \caption{Plot of the orientational correlation function $C_{S}(t)
      = \langle \cos\theta(t)\rangle$, where $\theta$ is the angle
      between the dumbbell's initial orientation and its orientation
      at time $t$, against $t$ for a single dumbbell in
      solution. Parameters are indicated in the legend.}
    \label{fig:S_t}
  \end{center}
\end{figure}

The orientational dynamics of the dumbbell molecules can be
characterized by the time $t_r$ it takes the orientational correlation
function, $C_{S}(t) = \langle \hat{{\bf e}}(t) \cdot \hat{{\bf e}}(0)
\rangle=\langle \cos\theta(t)\rangle$, to decay to 1/e of its initial
value. Here $\theta$ is the angle between the dumbbell's initial
orientation $\hat{{\bf e}}(0)$ and its orientation $\hat{{\bf e}}(t)$
at time $t$. This function is plotted in Fig.~\ref{fig:S_t} for both
inactive and active dumbbells. For inactive dumbbells $t_r \approx
300$. When the dumbbells are active, $t_r$ is shorter, particularly
for small $t_{c}$ and large $k$, with $t_r \approx 10$ for $t_{c}=20$
and $k=90$, and $t_r \approx 40$ for $t_{c}=200$ and $k=20$.

\begin{figure}[htbp]
  \begin{center}
    \includegraphics[height=1.0\columnwidth,angle=270]{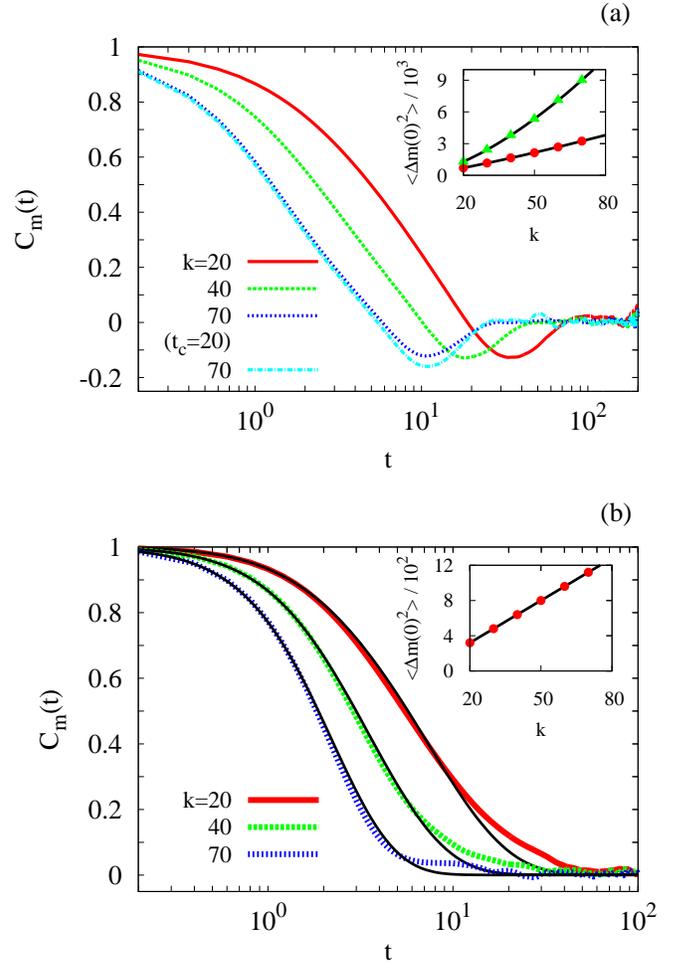}
    \caption{(a) Plot of $C_m(t)= \langle \Delta m(t)\Delta
      m(0)\rangle / \langle \Delta m^2 \rangle$, with $m =
      -k_0\ell(\ell-\ell_{0})$ the force dipole for a dumbbell,
      against $t$ for an active dumbbell with $t_{c}=100$. The value
      of $k$ is indicated in the legend. Also shown is data for
      $t_{c}=20$ at $k=70$. The value of $\langle \Delta m^{2}\rangle$
      is plotted in the inset as a function of $k$ for $t_{c}=100$
      (red circles) and $t_{c}=20$ (green triangles). The solid lines
      show the dependence of this quantity on $k^{\alpha}$ with
      $\alpha$ indicated in the plot. (b) Plot of $C_m(t)$ against $t$
      for an inactive dumbbell. Solid black lines show the theoretical
      prediction given by Eq.~(\ref{eq:langevin2}). The value of
      $\langle \Delta m^{2}\rangle$ is shown in the inset as a
      function of $k$, where the solid line shows a $k^{\alpha}$
      dependence with $\alpha=1$. }
    \label{fig:mcf_t}
  \end{center}
\end{figure}

The force dipole for a dumbbell molecule is $m(t) =
-k_0\ell(t)(\ell(t)-\ell_{0})$, where $k_0$ and $\ell_{0}$ stand for,
respectively, the spring constant and bond rest length at the time
when $m(t)$ is measured. We define the normalized temporal
force-dipole autocorrelation function by $C_m(t)=\langle \Delta
m(t)\Delta m(0) \rangle / \langle \Delta m^2 \rangle$, with $\Delta m
=m -\langle m \rangle$. It has an initial value of unity and decays to
zero at long times, since the asymptotic value of $\langle m(t)m(0)
\rangle$ is $\langle m \rangle^2$. This correlation function is
plotted in Fig.~\ref{fig:mcf_t}(a) for several values of the force
constant $k$.  The force dipole correlations decay with a strongly
damped oscillatory tail at longer times that is due to the changes in
sign when the forces that trigger closing or opening change their
sign. The force dipole correlation time $t_m$, defined as the time for
$C_m(t)$ to decay to 1/e of its initial value, ranges from $t_m
\approx 2-7$ for the data in the figure. These times are less than an
order of magnitude shorter than the orientational correlation times.
For active dumbbells $\langle \Delta m^{2}\rangle$ depends on $k$,
$\ell_{o,c}$ and $t_{c}$, and its magnitude decreases with increasing
$t_{c}$.  We find that it scales with the force constant as $\langle
\Delta m^{2}\rangle \sim k^{\alpha}$, where $\alpha$ also changes with
$t_{c}$. For the results shown in the inset to
Fig.~\ref{fig:mcf_t}(a), we find $\alpha \sim 1.6$ for $t_{c}=20$ and
$\alpha\sim1.2$ for $t_{c}=100$. In the limit $t_{c} \to \infty$,
$\alpha=1$.

The results for an active dumbbell may be contrasted with those for an
inactive dumbbell that simply experiences thermal fluctuations about
its open conformation. The correlation function $C_m(t)$ for this
situation is plotted in Fig.~\ref{fig:mcf_t}(b). It decays
monotonically to its long-time value, signalling the absence of
anti-correlation effects that arise from the active dumbbell
conformational changes. In addition, $\langle \Delta m^{2}\rangle$ now
scales as $\langle \Delta m^{2}\rangle \sim k$.

A simple Langevin model,
\begin{equation}
\label{eq:langevin} \mu \frac{d^2 \ell(t)}{dt^2}= -\zeta \frac{d
\ell(t)}{dt} -\mu \omega_o^2 (\ell(t)-\ell_o) +f(t),
\end{equation}
can be used to compute $C_m(t)$ for an inactive dumbbell. In this
equation $\zeta$ is the friction coefficient, $\mu=m_b/2$ is the
relative dumbbell mass, $\omega_o^2=k_o/\mu$ and $f(t)$ is a Gaussian
white-noise random force with correlation function $\langle
f(t)f\rangle= 2 k_B T\zeta \delta(t)$. The force dipole here takes the
form $m(t)=-k_o \ell(t)(\ell(t)-\ell_o)$ with $k_o=k$.  Using the
solution of Eq.~(\ref{eq:langevin}), the unnormalized force dipole
correlation function is given by
\begin{eqnarray}
\label{eq:langevin2} &&\langle m(t) m(0)\rangle =(k_BT)^2 \Big\{1 + \nonumber \\ &&
2e^{-\gamma t}\big( 1+\big(2 + \big(\frac{\omega_o}{\omega} \big)^2\big) \sinh^2
\omega t + \frac{\gamma}{2\omega} \sinh 2\omega t\big) + \nonumber \\ && \frac{k \ell_o^2}{k_B T} e^{-\gamma t/2
}\big( \frac{\gamma}{2 \omega} \sinh \omega t +\cosh \omega t
\big)\Big\},
\end{eqnarray}
where $\gamma=\zeta/\mu$, $\omega=\sqrt{\gamma^2/4- \omega_o^2}$ with
$\gamma > 2 \omega_o$ for overdamped dynamics. Its limiting form is
$\lim_{t \to \infty}\langle m(t) m(0)\rangle= \langle
m\rangle^2=(k_BT)^2$, which we may use to calculate $C_m(t)$. The only
unknown parameter in the expression for $\langle m(t) m(0)\rangle$ is
the friction coefficient $\zeta$ that appears in the ratio
$\gamma=\zeta/\mu$. By fitting to the data for a single force
constant, we obtain a value of $\gamma\sim3.19$, from which one
obtains good agreement with the simulation results for all values of
the force constant shown in Fig.~\ref{fig:mcf_t}(b). In the strongly
overdamped limit Eq.~(\ref{eq:langevin2}) takes the simpler form,
\begin{equation}
\langle m(t) m(0)\rangle =(k_BT)^2 \Big\{1 + 2 e^{-2kt/\zeta} + \frac{k \ell_o^2}{k_B T} e^{-kt/\zeta}\Big\}.
\end{equation}

For inactive dumbbells $ \langle \Delta
m^2\rangle=(k_BT)^2\left(2+k\ell_o^2/k_BT \right)$, which shows the
linear scaling of $ \langle \Delta m^2\rangle$ with $k$ seen in the
inset to Fig.~\ref{fig:mcf_t}(b). Since $k \ell_o^2/2k_BT \gg 1$, to a
good approximation we may write $ \langle m^2\rangle/k \ell_o^2
\approx k_BT$ and this ratio is approximately independent of the force
constant magnitude. The decay from this initial value of $C_m(t)$
depends on the value of the $k$, with a larger value resulting in a
faster decay, as can be seen in Fig.~\ref{fig:mcf_t}(b). The decay
time $t_m$ varies between $t_m \approx 2-8$, comparable to that for
active dumbbells, albeit from a smaller initial value.

\section{Diffusion in a field of active dumbbells}
\label{sec:diffusion}

\begin{figure}[htbp]
  \begin{center}
    \includegraphics[height=0.8\columnwidth,angle=0]{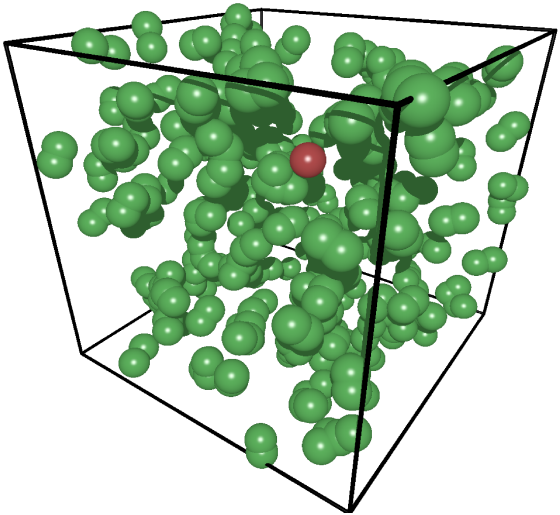}
    \caption{Instantaneous configuration of the system showing active
      dumbbells (green) and the single passive particle (red) for a
      system with dumbbell volume fraction $\phi=0.133$. }
    \label{fig:diagrams}
  \end{center}
\end{figure}

The diffusion of a passive particle as well as the self-diffusion of
an active dumbbell, in a field of active dumbbell molecules, are
discussed in this section.  A visual representation of the system
under study is given in Fig.~\ref{fig:diagrams}, which shows an
instantaneous configuration of the dumbbells and passive particle
drawn from the dynamics. Solvent particles are not displayed due to
their large number.

All our simulations start from an isotropic configuration of dumbbell
particles, and for all the system parameters studied here we find no
evidence of positional or orientational ordering of the dumbbells.

\subsection*{Passive particle diffusion}
As briefly described in the Introduction, the diffusion coefficients
of passive particles are enhanced when the medium in which they move
contains active enzymes or swimmers. Recent experimental studies have
shown that even on microscopic scales the diffusion of passive
molecular tracers are enhanced in the presence of active catalyst
molecules.~\cite{ref:Sen-tracer} In this subsection, we determine how
the diffusion coefficient of a passive tracer particle varies as a
function of the volume fraction of active dumbbells. Although our
study is motivated by the diffusive dynamics of active enzyme systems,
no specific enzymatic system is considered. Rather, we explore the
dependence of the diffusion coefficients on a wide range of dumbbell
and other system parameters.

Consider a single passive particle immersed in a system of volume $V$
containing $n_{\mathrm{db}}$ dumbbells with volume fraction
$\phi=v_{\mathrm{db}}n_{\mathrm{db}}/V=v_{\mathrm{db}}c$, where
$v_{\mathrm{db}}$ is the volume of a dumbbell and
$c=n_{\mathrm{db}}/V$ is the dumbbell concentration.  We define this
volume to be that of overlapping monomer spheres with radius
$\sigma_c/2$ in the open configuration,
$v_{\mathrm{db}}=2\pi[\sigma^{3}_{c}/2-(\sigma_{c}-l_{o})^2(2\sigma_{c}+l_{o})/8]/3$
with $v_{\mathrm{db}}=82.96$ for the parameters used here. The
effective volume will also vary as the dumbbell undergoes
conformational changes, so the volume fraction obtained using this
value of $v_{\mathrm{db}}$ simply provides a convenient way to specify
the dumbbell concentration, $c$.

The diffusion coefficient $D$ may be determined from the long-time
limit of the mean square displacement~\footnote{As the volume fraction
  increases the mean square displacement will generally exhibit
  subdiffusive dynamics on intermediate time scales. Here we focus on
  the diffusion coefficient determined in the long-time regime where
  normal diffusion again observed.}  $\Delta R^2(t)= \langle |{\bf
  R}(t)-{\bf R}(0)|^2 \rangle \sim 6D t$, where ${\bf R}$ is the
position of the passive particle and the angular brackets denote a
time and ensemble average.  In general the diffusion coefficient will
depend on the dumbbell volume fraction, force constants and average
hold time, $D=D(\phi,k,t_c)$.  In the absence of dumbbells ($\phi=0$)
it will be denoted by $D_0$ and for our system parameters this has the
value $D_0=1.14 \times 10^{-3}$. The thermal diffusion coefficient of
the passive particle in a solution containing inactive dumbbells,
denoted by $D_T(\phi)$, will also be a function of the dumbbell volume
fraction, since crowding by dumbbells will alter its value.

\begin{figure}[tbp]
  \begin{center}
   \includegraphics[height=\columnwidth,angle=270]{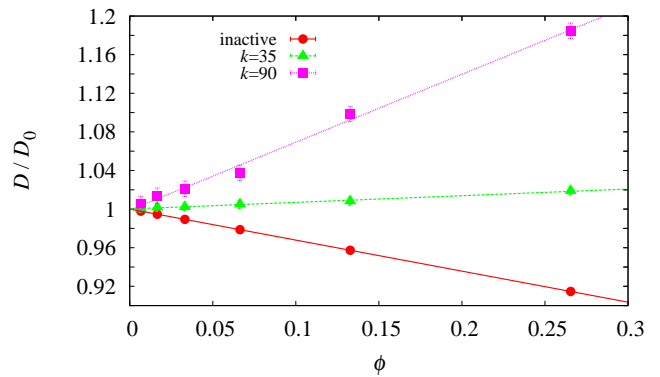}
   \caption{Diffusion coefficient $D$ of the passive particle versus
     dumbbell volume fraction $\phi$, in a system of active dumbbells
     with $t_{c}=20$ and two values of the force constant $k$
     indicated in the figure. Also shown is $D_T$, the diffusion
     coefficient of a passive particle in a system of inactive
     dumbbells. Data is normalized by $D_0$, the diffusion coefficient
     of a single passive particle in the absence of dumbbells.}
    \label{fig:D_c}
  \end{center}
\end{figure}
Figure~\ref{fig:D_c} compares the dependence of $D$, for two values of
$k$ and fixed $t_c=20$, and $D_T$ on $\phi$. (Additional data is given
in the Appendix B.)  For inactive dumbbells $D_T(\phi)$ decreases with
increasing $\phi$, consistent with the fact that the inactive
dumbbells act as crowding agents and inhibit the diffusive dynamics of
the passive particle.  The solid line in the figure shows that the
diffusion coefficient varies linearly with $\phi$ over the range of
volume fractions presented. Expressed as a function of $\phi$, we have
$D_T(\phi)=D_0(1+\kappa_o \phi)$, where the constant
$\kappa_o=-0.321$, which is independent of $k$, is obtained from a fit
to the data. The dependence on the dumbbell volume fraction is
different when the dumbbells are active: now $D$ increases with
increasing $\phi$ over the same range of $\phi$ values, rather than
decreasing as is the case for crowding by inactive dumbbells. We may
write $D(\phi,k,t_c)=D_0(1+\kappa(k,t_c) \phi)$, where $\kappa$
depends on $k$ and $t_c$, with $\kappa=0.701$ and 0.067 for $k=90$ and
$35$, respectively, for the data in Fig.~\ref{fig:D_c}.

\begin{figure}[htbp]
  \begin{center}
    \includegraphics[height=1.0\columnwidth,angle=270]{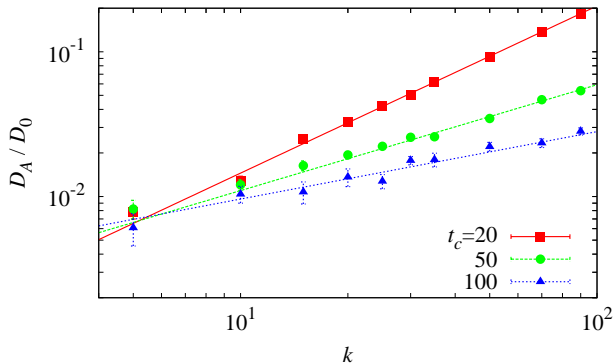}
    \caption{Active contribution to the diffusion coefficient,
      $D_{A}/D_0$, as a function of $k$ for several values of $t_{c}$
      and $\phi=0.133$. Points show simulation data, lines show fits
      of the form $D_{A}/D_0 = \lambda k^{\delta}\phi$ to the data.
            }
    \label{fig:D_k_tc}
  \end{center}
\end{figure}
The active contribution to the diffusion coefficient, $D_A$, is
defined by the equation $D= D_T+D_A$. Extensive computations of
$D_A(\phi,k,t_c)$ have been carried out to determine its dependence on
the dumbbell volume fraction, force constant and average hold
time. Figure~\ref{fig:D_k_tc} shows that for fixed $\phi$, $D_A$ has a
power-law dependence on $k$ of the form $D_A \sim k^{\delta}$, where
the exponent $\delta$ depends on $t_c$.  For a fixed value of $k$, as
$t_{c}$ increases $D_{A}$ decreases. The larger $t_c$ is, the slower
$D_A$ increases with $k$. The results of these simulations may be
summarized in the following form for the diffusion coefficient:
\begin{equation}
\label{eq:D_l_phi}
D(\phi,k,t_c)= D_T(\phi) +D_0\kappa_A(k,t_c)\phi,
\end{equation}
where $\kappa_A(k,t_c)=\kappa(k,t_c)-\kappa_o=\lambda(t_c)
k^{\delta(t_c)}$. Additional information on the dependence of
$\lambda$ and $\delta$ on $t_c$ is given in Appendix B. These results
are applicable provided $k$ is not too small since as $k \to 0$ the
dumbbell bond will soften and the dumbbell will dissociate. Note that
as $t_c \to \infty$ we have $\kappa(k,t_c) \to \kappa_c$, its value
for a dumbbell that fluctuates about its closed conformation.

Depending on the system parameters, it is possible that the decrease
in diffusion due to crowding-induced hindered motion may be larger
than the increase due to dumbbell activity. In such a circumstance
$D(\phi)/D_0$ may be less than unity ($\kappa (k,t_c)<0$), although
the ratio will be larger than $D_T(\phi)/D_0$ for systems with
inactive dumbbells. In our simulations we found that $\kappa(k,t_c)
>0$ for most values of $k$ and $t_c$, although for a given $t_c$ there
is a $k$ value at which $\kappa(k,t_c)$ will change sign. This will
occur when $\kappa(k,t_c)=0$, which corresponds to
$k=(-\kappa_o/\lambda(t_c))^{1/\delta(t_c)}$, given the scaling forms
below Eq.~(\ref{eq:D_l_phi}).

The magnitude of the enhancement of the diffusion coefficient as
measured by $D_A/D_T$ depends strongly on the system parameters,
$\phi$, $k$ and $t_c$, as well as $\tau_{{\text MPC}}$, which
determines the solvent properties. The largest enhancements for
$k=90$, $t_c=20$, $\tau_{{\text MPC}}=0.01$ in Fig.~\ref{fig:D_c} are
$D_A/D_T \approx 0.15$ and $D_A/D_T \approx 0.3$ for $\phi=0.133$ and
$0.266$, respectively. For another set of system parameters with
$\tau_{{\text MPC}}=0.05$, corresponding to a smaller solvent
viscosity, $t_c=20$ and $k=9$, we find $D=4.68 \times 10^{-3}$ and
$D_0=4.44 \times 10^{-3}$ for $\phi=0.133$ giving $D_A/D_T \approx
0.05$. In addition to the quantitative estimates of the diffusion
enhancement, several qualitative features of $D_A$ are worth
summarizing. The coefficient $\kappa_A$ differs from $\kappa_o$ since
it depends strongly on $k$ and $t_c$.  For fixed $t_c$, $\kappa_A \sim
k^\delta$ where the exponent $\delta$ decreases with increasing
$t_c$. These qualitative features of $D_A$ differ markedly from those
that characterize the behavior of $D_T$.

Active contributions to passive particle diffusion from the collective
hydrodynamic interactions of many active proteins were discussed
earlier using a Langevin model.~\cite{ref:Mikhailov15} The result was
the following estimate for $D_A$:
\begin{equation}\label{eq:D_A_th}
D_{A}^{\text{th}}= \frac{S_{A}}{60\pi\ell_{{\text cut}}\eta^{2}v_{\mathrm{ex}}}\phi\equiv D_0\kappa_A^{\text{th}}(k,t_c) \phi,
\end{equation}
for a uniform distribution of proteins with concentration
$c=\phi/v_{\mathrm{ex}}$. This order-of-magnitude estimate was derived
assuming a random static distribution of protein orientations, slow
protein translational dynamics, and Oseen interactions with a
short-distance cut-off, $\ell_{{\text cut}}$, taken to be the sum of
the effective radii of the passive particle and protein.  In this
equation $S_A$ characterizes the strength of the force dipole
correlations, $\langle\Delta m(t) \Delta m(0)\rangle $. The last
equality defines the theoretical estimate of $\kappa_A$. Evaluation of
Eq.~(\ref{eq:D_A_th}) for a range of $k$ and $t_c$ values shows that
$\kappa_A^{\text{th}}(k,t_c) \sim k^{\sigma}$ for fixed $t_c$, where
$\sigma < \delta$, smaller than the exponent $\delta$ found from
simulation. The predicted values of the diffusion enhancement are
consistent with Eq.~(\ref{eq:D_A_th}) since they differ by less than
an order-of-magnitude from those in our microscopic simulations; for
example, for $\phi=0.133$, $k=35$ and $t_c=20$ we have
$D_{A}^{\text{th}}/D_T=0.01$ while from simulation $D_{A}/D_T=0.051$.

\subsection*{Active dumbbell self-diffusion}
\begin{figure}[htbp]
  \begin{center}
    \includegraphics[height=0.9\columnwidth,angle=270]{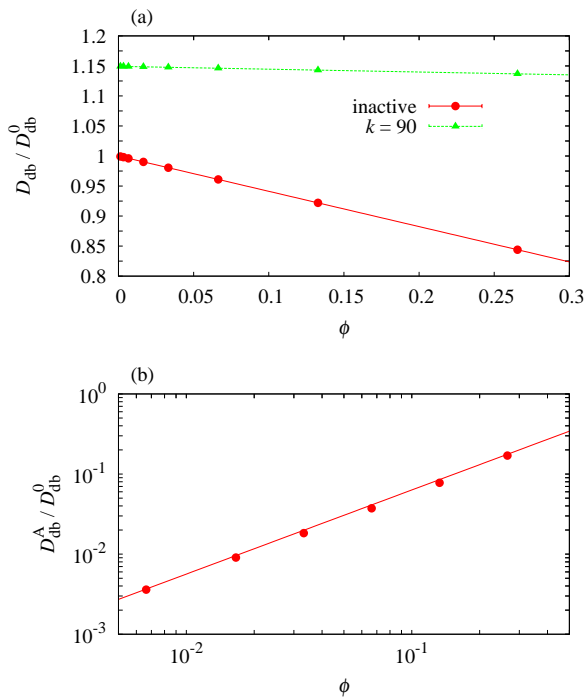}
    \caption{(a) Dumbbell self-diffusion coefficient $D_{\mathrm{db}}$
      versus $\phi$, for a system of active dumbbells with $t_{c}=20$
      and $k=90$ (green triangles), and $D_{\mathrm{db}}^T$ for a
      system of inactive dumbbells (red circles). Data is normalized
      by $D_{\mathrm{db}}^0$, the diffusion coefficient of a single
      inactive dumbbell. (b) Active contribution to the dumbbell
      self-diffusion coefficient $D^A_{\mathrm{db}}$ versus $\phi$.}
    \label{fig:D_c_db}
  \end{center}
\end{figure}

The self-diffusion coefficients of the dumbbells themselves,
$D_{\mathrm{db}}$, are also modified due to crowding by other
dumbbells, and the effects of crowding differ depending on whether the
dumbbells are active or inactive. Figure~\ref{fig:D_c_db}(a) shows
$D_{\mathrm{db}}^T$ and $D_{\mathrm{db}}$, the diffusion coefficients
for inactive and active dumbbells, respectively, as a function of
$\phi$, normalized by the diffusion coefficient of a single inactive
dumbbell, $D_{\mathrm{db}}^0 = 1.625 \times 10^{-3}$. For inactive
dumbbells $D_{\mathrm{db}}^T$ decreases with increasing $\phi$, as
expected for a crowded environment.  A similar trend is seen for
active dumbbells, with $D_{\mathrm{db}}$ also decreasing with an
increase in $\phi$, although the decrease is much smaller than that
for inactive dumbbells. As discussed above, recall that even for the
passive particle in a system of active dumbbells, depending on the
system parameters, the diffusion coefficient $D$ may decrease as
$\phi$ is increased, but this decrease will be less strong than that
for a system of inactive dumbbells. In these cases the effects of
activity are not sufficient to completely overcome the tendency for
crowding to decrease the diffusion coefficient. Nevertheless,
$D_{\mathrm{db}}/D_{\mathrm{db}}^T> 1$ and the magnitudes of the
diffusion coefficient changes are comparable to those for the passive
particle. For example, for a volume fraction of $\phi=0.133$ and
$k=90$, $t_c=20$, $\tau_{{\text MPC}}=0.01$ we have
$D_{\mathrm{db}}/D_{\mathrm{db}}^T \approx 1.19$, while for $k=9$,
$t_c=20$, $\tau_{{\text MPC}}=0.05$ we have $D_{\mathrm{db}}/D_{{\text
    db}}^T \approx 1.05$.

Although the tendency for $D_{\mathrm{db}}$ to increases or decrease
with increasing $\phi$ depends on the system parameters, one can see
that the values of the dumbbell self-diffusion coefficients differ
markedly depending on whether they are active or inactive. In contrast
to $D$ for passive particle diffusion, which tends to a common value
of $D_0$ as $\phi \to 0$ regardless of whether the dumbbells are
active or inactive, even for a single dumbbell in solution
$D_{\mathrm{db}}$ will be different if it is active or inactive. For
example, for the data in Fig.~\ref{fig:D_c_db}(a),
$D_{\mathrm{db}}(\phi=0)=2.0 \times 10^{-3}$ and $D_{{\text
    db}}/D_{\mathrm{db}}^0 \approx 1.15$ for a single dumbbell in
solution. Since $D_{\mathrm{db}}$ depends on the dumbbell conformation
and is larger when the dumbbell is in the compact closed form, one
expects, and finds, $D_{\mathrm{db}}> D_{{\text db}}^T$.~\cite{[{Such
    dependence of diffusion on conformational dynamics was observed
    earlier in an investigation of active enzyme dynamics for
    adenylate kinase (AKE).  The enzyme catalyzes the reversible
    reaction $AMP + ATP \rightleftharpoons 2ADP$ and in the process
    changes its conformation and cycles from open to closed forms. The
    diffusion coefficient for the catalytically active enzyme was
    found to be larger that for the enzyme is in its open conformation
    when no substrate is bound, $D_{{\text AKE}}/D^T_{{\text
        AKE}}\approx 1.17$, similar to the results for the dumbbell
    model considered here. }] carlos2011} Furthermore, when the
average hold time $t_{c}$ is smaller, $D_{\mathrm{db}}(\phi=0)$
becomes smaller, as the dumbbell spends less time in its closed
conformation. This results in a measured $D_{\mathrm{db}}$ that is
smaller at low $k$ values for dumbbells with shorter average hold
times than for ones with larger $t_{c}$ values, as can be seen in
Table~\ref{tab:3} in Appendix B.

Accounting for the fact that $D_{\mathrm{db}}$ is different for single
active and inactive dumbbells in solution, we define the active
dumbbell contribution to the self-diffusion coefficient by the
equation, $D_{\mathrm{db}}=D^T_{\mathrm{db}}+D^A_{{\text
    db}}+D^{A,0}_{\mathrm{db}}$, where
$D^{A,0}_{\mathrm{db}}=D_{\text{
    db}}(\phi=0)-D^0_{\mathrm{db}}$. Figure~\ref{fig:D_c_db}(b) plots
$D^{A}_{\mathrm{db}}$ versus $\phi$ and shows that the active
contribution increases with increasing $\phi$.

The results presented above show that hydrodynamic interactions
resulting from nonequilibrium force dipole fluctuations of the
dumbbell molecules give rise to enhanced diffusion of both the passive
particle and the dumbbells themselves when compared to their values
for systems containing only inactive dumbbells. Direct intermolecular
interactions also play a role but estimates based on dumbbell sizes
and ranges of intermolecular forces suggest that these interactions
are important only at the highest volume fractions considered in this
study. Contributions from direct intermolecular interactions will
increase in importance as the average separation between dumbbells
$l_{\mathrm{sep}}=(v_{\mathrm{ex}}/\phi)^{1/3}$ approaches the maximum
length of a dumbbell $l_{\mathrm{db}}=\ell_{o} + 2 \sigma_{c}
=8.3$. This is the case only for systems with the largest two volume
fractions studied, $\phi=0.133$ and $\phi=0.266$, where
$l_{\mathrm{sep}}=8.54$ and $6.78$, respectively.

\section{Conclusions}\label{sec:Conclusions}
The microscopic simulation systems containing a passive particle and
either active or passive dumbbell molecules has allowed us to explore
how diffusive dynamics varies with dumbbell activity and volume
fraction.  The results showed that the diffusive dynamics of passive
particles in systems crowded by active molecules that change their
conformations differs markedly from that when the crowding molecules
are inactive. The self-diffusion coefficients of the crowding
molecules themselves also display properties that depend on their
activity. While crowding by molecules that thermally fluctuate about
their open (or closed) metastable states leads to well-known
subdiffusive dynamics and diffusion coefficients that decrease with
increasing volume fraction, diffusion coefficients are enhanced, or
decrease more slowly, when the crowding agents are active.

Hydrodynamic interactions induced by active force dipole fluctuations
are responsible for the observed diffusion coefficient increases, and
direct intermolecular interactions contribute at the highest volume
fractions. The particle-based dynamical model used in this
investigation accounts for both of these effects and permits a
detailed analysis of the phenomena. The magnitudes of the changes to
the diffusion coefficient were shown to depend not only on the volume
fraction of dumbbells, but also on the force dipole strength and the
mean times spent in the open or closed conformations. The diffusion
enhancement in experiments and in our simulations is not large but its
existence signals that conformational changes arising from catalytic
activity play a role in transport in active systems.

The dumbbell conformational changes in this study were specified by a
stochastic model that was chosen to mimic some features of the cyclic
dynamics of enzymes undergoing conformational changes during their
catalytic operation. Our dynamical model can be generalized to include
a more detailed description of the substrate binding, unbinding and
reaction processes with the enzyme so that the dependence of the
effects on substrate concentration can be investigated. More realistic
enzyme models may also be employed.~\cite{carlos2011,inder2014} The
results in this paper provide the basis for the development and
further study of more realistic models to probe transport properties
in systems crowded by chemically active and inactive molecules and
their relevance to biochemical processes in the cell.

\vspace{0.1in} \noindent Acknowledgements: We would like to thank
Alexander Mikhailov for useful discussions on this topic.  The
research of RK was supported in part by a grant from the Natural
Sciences and Engineering Research Council of Canada.  MD thanks the
Humboldt foundation for financial support. RK and MD were partially
supported through the research training group GRK 1558 funded by
Deutsche Forschungsgemeinschaft.

\section*{Appendix A}
\label{sec:appendixA}

Multiparticle collisions were implemented using the MPC-AT+a rule that
employs the Anderson thermostat and conserves linear and angular
momentum.~\cite{ref:cell1,*ref:noguchi2007,*ref:noguchi2007b} Allowing
the solvent particles to interact and exchange momentum only in the
collision step, the frequency of which can be chosen according to the
desired properties of the solvent, makes the method computationally
efficient. Since the collision rule conserves mass and momentum at the
cell level, the hydrodynamic flow fields will be described correctly,
a feature that is essential for hydrodynamic interactions. The solvent
viscosity $\eta$ can be controlled by varying the number of solvent
particles $n_{f}$ per collision cell and/or the MPC time,
$\tau_{{\text MPC}}$.~\cite{Kapral_08,ref:Gompper} In our simulations,
we use a cubic simulation box of linear size $n_{x}=50$ MPCD cells in
each direction, with $n_{f}=10$ fluid particles per cell, and an MD
time step of $\Delta t=0.001$.

The choice of parameters was based on several criteria. We wish to
study a large range of force constants in order to identify the
underlying trends in the system behaviour. Since inertia does not play
a significant role in protein dynamics in solution, the dumbbell
dynamics should be overdamped. This condition sets a limit on the
maximum value of the spring constant. Furthermore, for spring energies
which are comparable to the thermal energy ($k\sim k_{B}T/a^2$),
thermal fluctuations dominate the dumbbell motion. This lower bound
depends only on the value of $k_{B}T$, and not on any other fluid
parameters. We consider two values of the MPC time, $\tau_{{\text
    MPC}}=0.01$ which gives a large fluid viscosity of $\eta\sim 36$,
and $\tau_{{\text MPC}}=0.05$ which gives a lower fluid viscosity of
$\eta\sim 8.4$. In simulations with the higher viscosity a large range
of force constant values may be used while still remaining in the
regime of overdamped dynamics.

The dumbbell mass is chosen so that it is neutrally buoyant. Since the
dumbbell interacts with the solvent only in the MPC collision step, we
define the volume of interaction of a dumbbell with the fluid by
$v_{\mathrm{f,db}}=4\pi r_{\mathrm{f,db}}^{3}/3$, where
$r_{\mathrm{f,db}}=(\ell_{o}+\ell_{c})/2$, which corresponds to a
sphere with a diameter equal to the average dumbbell spring rest
length. Note that this is distinct from the dumbbell volume
$v_{\mathrm{db}}$ defined earlier. This then gives a total dumbbell
mass of $m_{\mathrm{db}}=v_{\mathrm{f,db}}n_{f}$. Throughout we set
$\ell_{c}=2$ and $\ell_{o}=4$, such that we have
$m_{\mathrm{db}}=141.37$. From our fluid and dumbbell parameters we
then obtain a cross-over spring constant between overdamped and
underdamped motion of $k_{o}> 90$ for $\tau_{{\text MPC}}=0.01$ and
$k_{o}>11$ for $\tau_{{\text MPC}}=0.05$.

To control the switching between the two spring rest lengths we must
set the cut-off parameter that the length must cross before the hold
time is chosen. We set $\delta\ell_{o}=\delta\ell_{c}=
0.05(\ell_{o}-\ell_{c})=0.1$. The hold times are chosen from log
normal distributions with averages for the open and closed
configurations of $t_{o}$ and $t_{c}$, and with scale parameter
$\sigma=0.5$. Throughout the time spent in the open configuration is
$t_{o}=0$.

For the passive particle we set the passive particle-fluid interaction
radius to be $\sigma_{cf}=2$ and the passive particle-dumbbell bead
interaction diameter to be $\sigma_{c}=4.30$. The passive particle is
also chosen to be neutrally buoyant, such that its mass is given by
$m_{c}=4\pi\sigma^{3}_{cf}n_{f}/3$ which for the parameters given here
is $m_{c}=335.103$.

Dimensionless units can be mapped approximately onto physical units by
matching dimensions and time scales.~\cite{ref:Padding} Consider
$\tau_{{\text MPC}}=0.05$ for which $D_0=4.44 \times 10^{-3}\;
a^2/t_0$. Taking a radius of 5 nm for the passive particle and
assuming $D_0$ is given by its Stokes-Einstein value one finds $t_0
\approx 0.1$ ns. For $k=9\; k_BT/a^2$ we then have for the forces
corresponding to active opening and closing $F \approx 20$ pN.

\section*{Appendix B}
\label{sec:appendixB}

This Appendix provides some of the numerical values of the data in the
plots, along with parameters that enter in the phenomenological forms
for the diffusion coefficients. The data used to construct
Fig.~\ref{fig:D_c} is in Table~\ref{tab:0}.
\begin{table}[tbp]
  \begin{center}
    \begin{tabularx}{8cm}{|C|C|C|C|}
      \hline
      $\phi/10^{-2}$ & $k$ & \shortstack{$D/10^{-3}$} & \shortstack{ $D_{T}/10^{-3}$} \\\hline\hline
      $0.66$  & $90$ & $1.145$ & $1.137$ \\
      $1.66$  & $90$ & $1.155$ & $1.134$ \\
      $3.32$  & $90$ & $1.163$ & $1.128$ \\
      $6.64$  & $90$ & $1.182$ & $1.115$ \\
      $13.3$ & $90$ & $1.261$ & $1.092$ \\
      $26.6$ & $90$ & $1.350$ & $1.043$ \\ \hline
      $0.66$  & $35$ & $1.140$ & $1.137$ \\
      $1.66$  & $35$ & $1.141$ & $1.134$ \\
      $3.32$  & $35$ & $1.142$ & $1.128$ \\
      $6.64$  & $35$ & $1.145$ & $1.115$ \\
      $13.3$ & $35$ & $1.148$ & $1.092$ \\
      $26.6$ & $35$ & $1.160$ & $1.043$ \\\hline
    \end{tabularx}
     \end{center}
  \caption{Values of $D$ and $D_{T}$ plotted in Fig.~\ref{fig:D_c} as
    functions of volume fraction $\phi$ for $t_c=20$ and two values of
    $k$, for systems with $\tau_{{\text MPC}}=0.01$.}
     \label{tab:0}
\end{table}

Table~\ref{tab:2a} gives representative values of the parameters that
enter in Eq.~(\ref{eq:D_l_phi}) for the passive particle diffusion
coefficient.
\begin{table}[htbp]
\begin{center}
\begin{tabularx}{8.5cm}{|C|C|C|C|C|C|} \hline
$\phi/10^{-2}$ & $t_c$ & $\lambda/10^{-3}$ & $\delta$ & \shortstack{Fit \\ $D_{T}/10^{-3}$} & \shortstack{Sim \\ $D_{T}/10^{-3}$} \\\hline\hline
$0.66$  & $20$  & $5.18$   & $1.13$ & $1.137$ & $1.137$ \\
$1.66$  & $20$  & $4.74$   & $1.12$ & $1.133$ & $1.134$ \\
$3.32$  & $20$  & $3.60$   & $1.16$ & $1.128$ & $1.128$ \\
$6.64$  & $20$  & $4.30$   & $1.12$ & $1.112$ & $1.115$ \\
$13.3$ & $20$  & $4.74$   & $1.16$ & $1.086$ & $1.092$ \\
$26.6$ & $20$  & $4.82$   & $1.18$ & $1.052$ & $1.043$ \\\hline
$13.3$ & $30$  & $11.0$  & $0.98$ & $1.087$ & $1.092$ \\\hline
$13.3$ & $50$  & $19.5$  & $0.73$ & $1.083$ & $1.092$ \\\hline
$13.3$ & $100$ & $37.1$  & $0.47$ & $1.084$ & $1.092$ \\\hline
$13.3$ & $200$ & $78.6$ & $0.20$ & $1.097$ & $1.092$ \\\hline
    \end{tabularx}
  \end{center}
  \caption{Values of $\lambda$, $\delta$ and $D_{T}$ found from
    fitting $D = D_{T} + D_0\lambda k^{\delta}\phi$ to the data, for
    systems with $\tau_{{\text MPC}}=0.01$, and with volume fraction and
    average hold time $t_c$ indicated. The last column gives $D_{T}$ as
    measued in a system of inactive dumbbells.}
  \label{tab:2a}
\end{table}

Table~\ref{tab:3} presents data for the dumbbell self-diffusion
coefficient for several values of the hold time $t_c$ and force
constant $k$.
\begin{table}[tbp]
  \begin{center}
    \begin{tabularx}{8cm}{|C|C|C|} \hline
$t_c$ & $k$ & $D_{\mathrm{db}}/10^{-3}$ \\\hline\hline
$30$ & $5$ & $1.707$ \\
$30$ & $10$ & $1.727$ \\
$30$ & $20$ & $1.758$ \\
$30$ & $35$ & $1.783$ \\
$30$ & $90$ & $1.908$ \\\hline
$50$ & $5$ & $1.717$ \\
$50$ & $10$ & $1.729$ \\
$50$ & $20$ & $1.743$ \\
$50$ & $35$ & $1.774$ \\
$50$ & $90$ & $1.847$ \\\hline
$100$ & $5$ & $1.724$ \\
$100$ & $10$ & $1.741$ \\
$100$ & $20$ & $1.753$ \\
$100$ & $35$ & $1.770$ \\
$100$ & $90$ & $1.817$ \\\hline
${\text{ inactive}}$ & $90$ & $1.625$ \\\hline
    \end{tabularx}
  \end{center}
  \caption{Diffusion coefficients $D_{\mathrm{db}}$ of dumbbell
    particles for systems with $\tau_{{\text MPC}}=0.01$, at average
    hold times and with spring constants indicated in the table. All
    other parameters are set to the values used throughout, given in
    the text of the paper. Data for both a system of active dumbbells
    with average hold time $t_{c}$ and a system of inactive dumbbells
    are presented, all at a volume faction of $\phi=0.133$.}
  \label{tab:3}
\end{table}

\bibliography{main}

\end{document}